\documentclass[a4pape,citesort]{PoS}
\usepackage{caption}
\usepackage{subcaption}
\usepackage{graphicx}
\usepackage{citesort}
\usepackage{bold-extra}
\usepackage{afterpage}
\usepackage[backend=bibtex,style=numeric,sorting=none,doi=false]{biblatex}
\def\ncteqpp{{\tt \textbf{nCTEQ++}}}
\def\wz{$W^\pm\!/Z$}

\usepackage{wrapfig}
\usepackage{lipsum,lineno}
\usepackage{microtype}  

\makeatletter 
\renewcommand\speaker[1]{\if@speaker\global\@dblspeaktrue\fi
			\global\@speakertrue
			\global\setbox\@firstaubox
			\hbox{{\let\thanks\@gobble
				\let\footnote\@gobble\small
				\rm  The nCTEQ Collaboration}}%
			#1\thanks{Speaker.}\
			}%
\makeatother

\title{nCTEQ PDFs at the LHC: 
 \hspace{\textwidth} 
{\it  Vector boson production in heavy ion collisions}
}

\ShortTitle{nCTEQ PDFs at the LHC}

\author{%
The  nCTEQ Collaboration:\thanks{%
We acknowledge the hospitality of CERN, DESY, and Fermilab where a
portion of this work was performed.
This work was also partially supported by the U.S.\ Department of
Energy under Grant No.\ DE-SC0010129.
}\qquad  \null  \hspace{\textwidth}
D.~B.~Clark\rlap,${}^1$ 
E.~Godat\rlap,${}^1$ 
T.~J.~Hobbs\rlap,${}^1$ 
T.~Je\v{z}o\rlap,${}^2$ 
J.~Kent\rlap,${}^1$ 
C.~Keppel\rlap,${}^3  $
M.~Klasen\rlap,${}^4$ 
K.~Kova\v{r}\'{i}k\rlap,${}^4$ 
A.~Kusina\rlap,${}^{5}$ 
F.~Lyonnet\rlap,${}^1$ 
J.G.~Morfin\rlap,${}^7$
F.~I.~Olness\rlap,${}^1$\speaker{}
J.F.~Owens\rlap,${}^8$
I.~Schienbein\rlap,${}^6$ 
J.~Y.~Yu${}^6$ 
\\
${}^1$Department of Physics, Southern Methodist University, Dallas, TX 75275, USA\\ 
${}^2$Physik-Institut, Universit\"at Z\"urich, Winterthurerstrasse 190, CH-8057 Z\"urich,
Switzerland\\
${}^3$Thomas Jefferson National Accelerator Facility, Newport News, VA, 23606, USA\\
${}^4$Institut f\"{u}r Theoretische Physik, Westf\"{a}lische Wilhelms-Universit\"{a}t
M\"{u}nster, \\ \qquad
Wilhelm-Klemm-Stra{\ss}e 9, D-48149 M\"{u}nster, Germany \\
${}^5$Institute of Nuclear Physics, Polish Academy of Sciences, \\ \qquad
ul. Radzikowskiego 152, 31-342 Cracow, Poland\\
${}^6$Laboratoire de Physique Subatomique et de Cosmologie, Universit\'{e}
Grenoble-Alpes, \\ \qquad
CNRS/IN2P3, 53 avenue des Martyrs, 38026 Grenoble,
France \\
${}^7$Fermi National Accelerator Laboratory, Batavia, Illinois 60510, USA\\
${}^8$Department of Physics, Florida State University, Tallahassee, Florida 32306-4350, USA\\
}

\abstract{
  Extraction of the strange quark PDF is a long-standing puzzle.
We use the nCTEQ nPDFs with uncertainties to study the impact of the
LHC $W/Z$ production data on both the flavor differentiation
and nuclear corrections; this complements the information from neutrino-DIS data.
As the proton flavor determination is dependent on nuclear corrections
(from heavy target DIS, for example), LHC heavy ion measurements can
also help improve proton PDFs.
We introduce a new implementation of the nCTEQ code (\ncteqpp) based
on C++ which has a modular strucure and enables us to easily integrate
programs such as HOPPET, APPLgrid, and MCFM.
Using ApplGrids generated from MCFM, we use \ncteqpp{} to perform a
preliminary fit including the $pPb$ LHC \wz{} vector boson data.
}

\FullConference{XXVII International Workshop on Deep-Inelastic Scattering and Related Subjects - DIS2019\\
		8-12 April, 2019\\
		Torino, Italy}

\begin{document}

\def\figdy{
\begin{wrapfigure}{R}{0.55\textwidth} 
\centering{} 
\vspace{-25pt}
\includegraphics[width=0.55\textwidth]{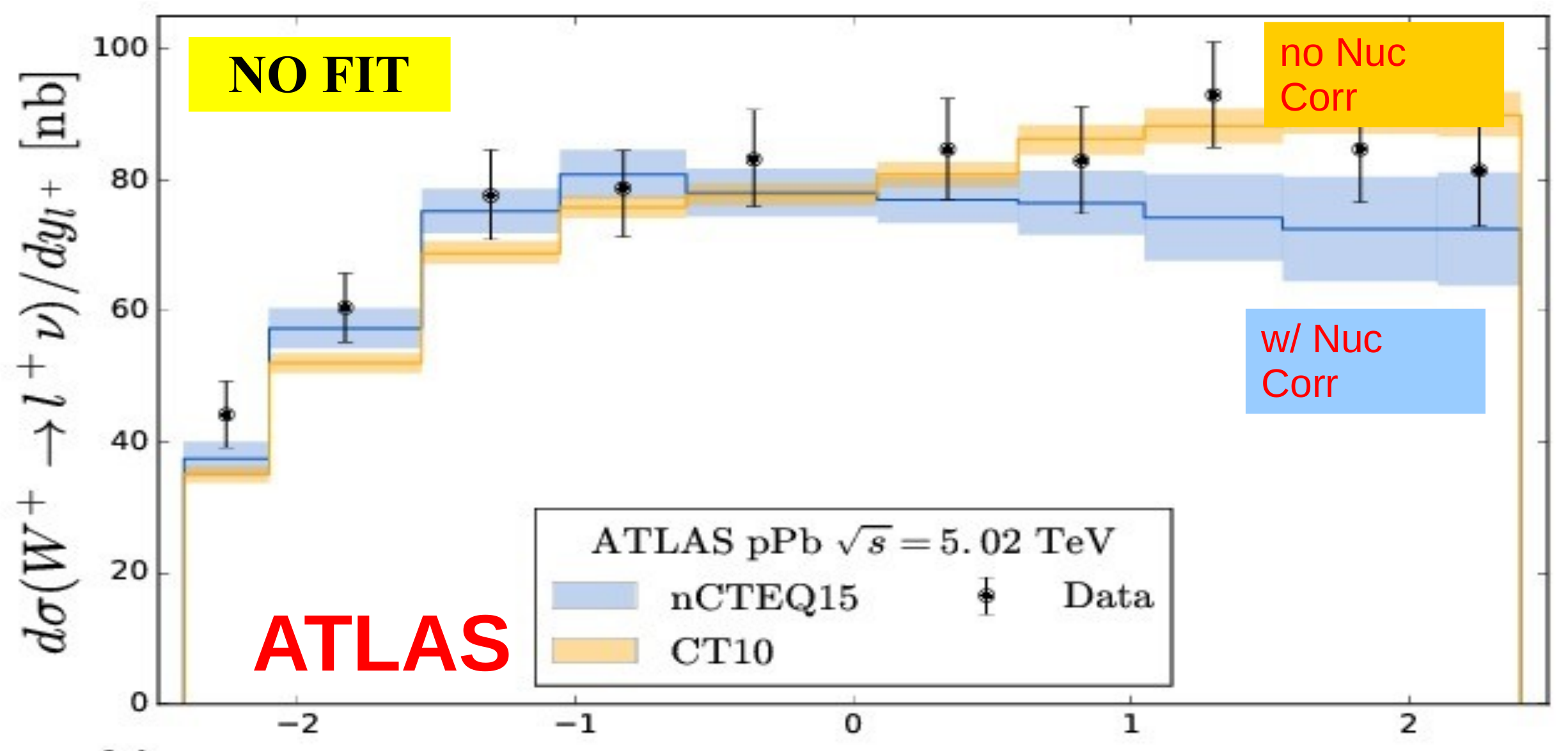}
\vspace{-20pt}
\caption{
Comparison of LHC $W$ boson production in $p$-$Pb$ processes {\it vs.} rapidity. 
The lighter (yellow) band uses CT10 with no nuclear corrections,
and the darker (blue) band uses the nCTEQ15 PDFs;
this data is {\it not} included in the nCTEQ15 fit~\cite{Kusina:2016fxy}.
}
\vspace{-10pt}
\label{fig:dsigdy}
\end{wrapfigure}
}

\def\figcomp{
\begin{figure}[b] 
\centering{} 
\includegraphics[width=0.45\textwidth]{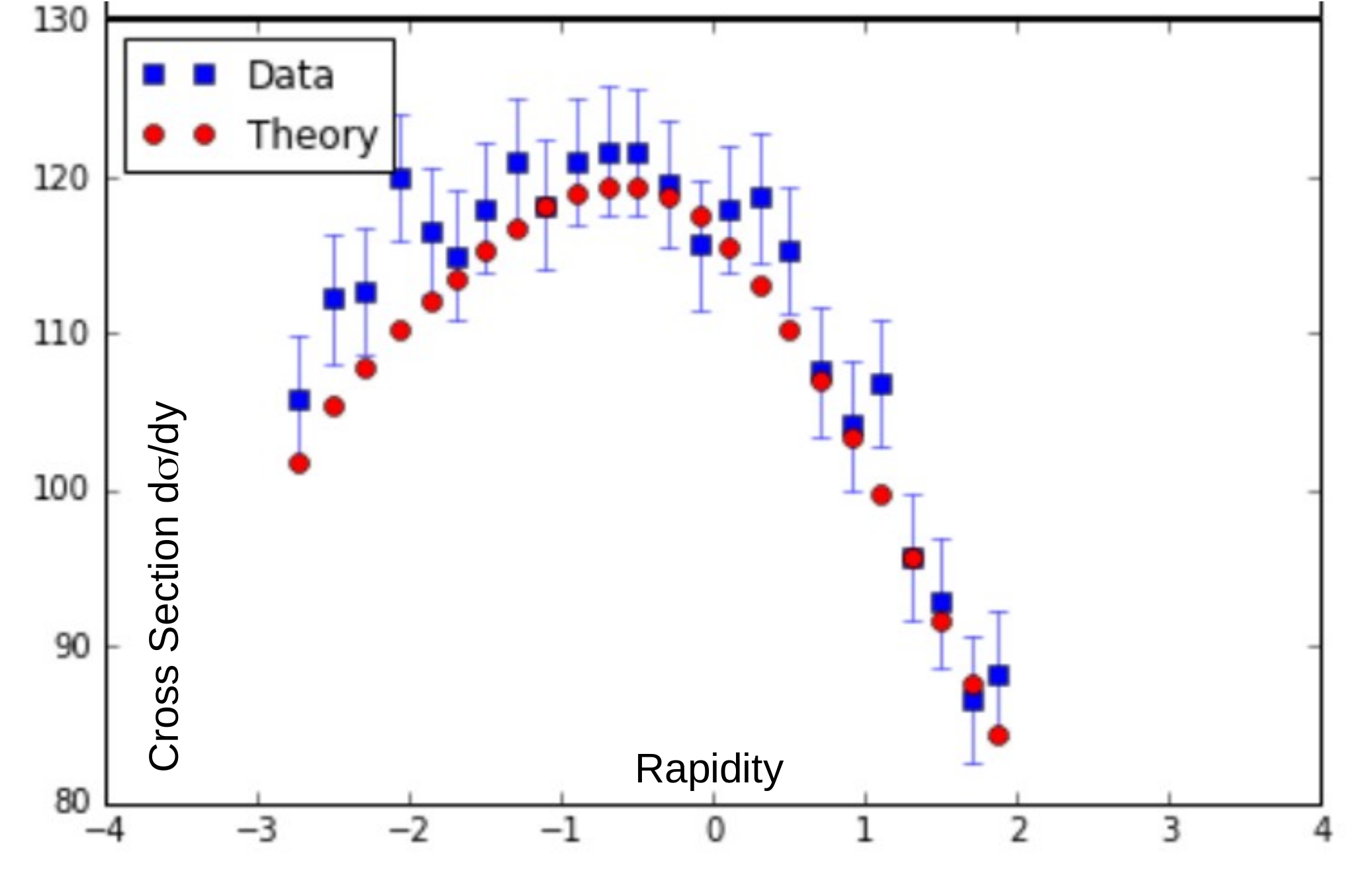}
\hfil
\includegraphics[width=0.48\textwidth]{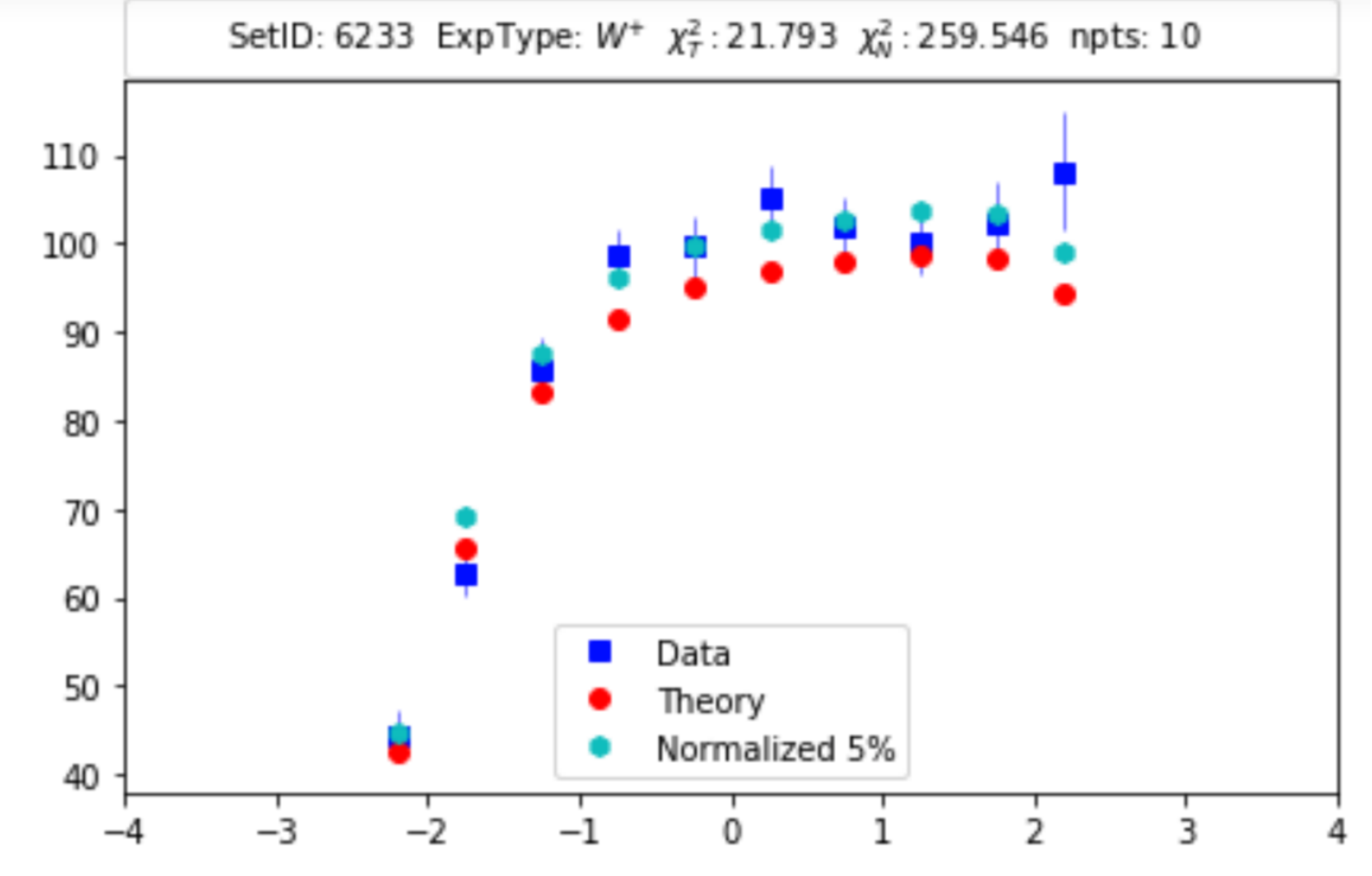}
\vspace{0pt}
\caption{ 
a)~Theory predictions for Run II CMS $W^-$ production (ID:6232), and 
b)~for Run I CMS $W^+$ production (ID:6233) in $pPb$.
The data are the blue squares and the theory are the red points.
For comparison, we also display the theory predictions in~b) with a 5\% normalization shift (cyan). 
}
\label{fig:comp}
\end{figure}
}

\def\figchi{
\begin{wrapfigure}{R}{0.65\textwidth} 
\centering{} 
\vspace{-15pt}
\includegraphics[width=0.65\textwidth]{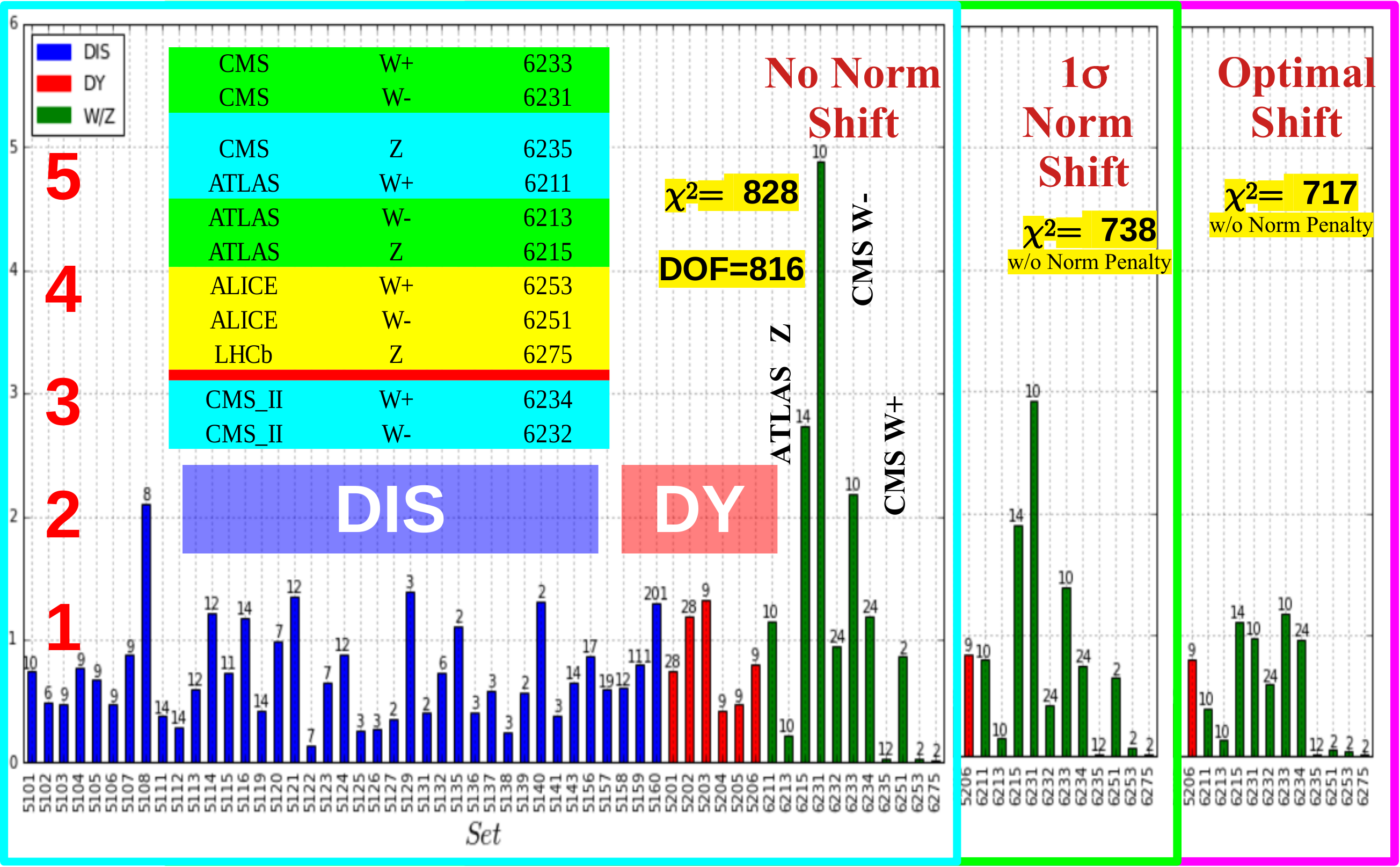}
\null \vspace{-10pt}
\caption{
$\chi^2/dof$ for individual data sets of the (restricted) nCTEQ+LHC fit;
data set ID's are given in  Ref.~\cite{Kusina:2016fxy}.
The  LHC \wz{} data is displayed in green.
Prelimnary results are shown for i)~no normalization shift,
ii)~a shift of up to $1\sigma$,
iii)~an~unconstrainted (optimal) shift. 
}
\label{fig:chi2}
\end{wrapfigure}
}

\def\figdouble{
\begin{figure}[tb] 
\centering{} 
\includegraphics[width=0.48\textwidth]{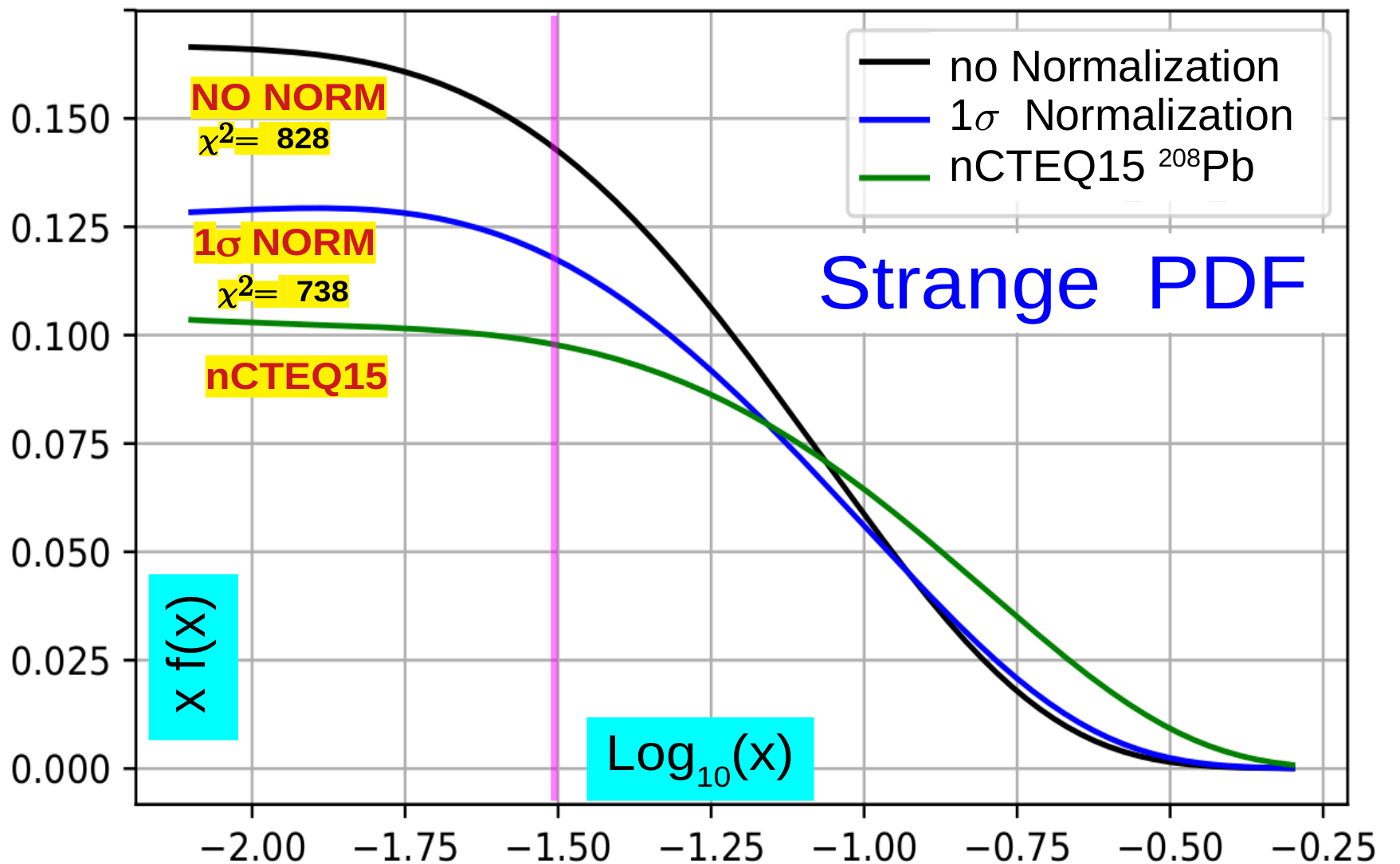}
\hfil
\includegraphics[width=0.48\textwidth]{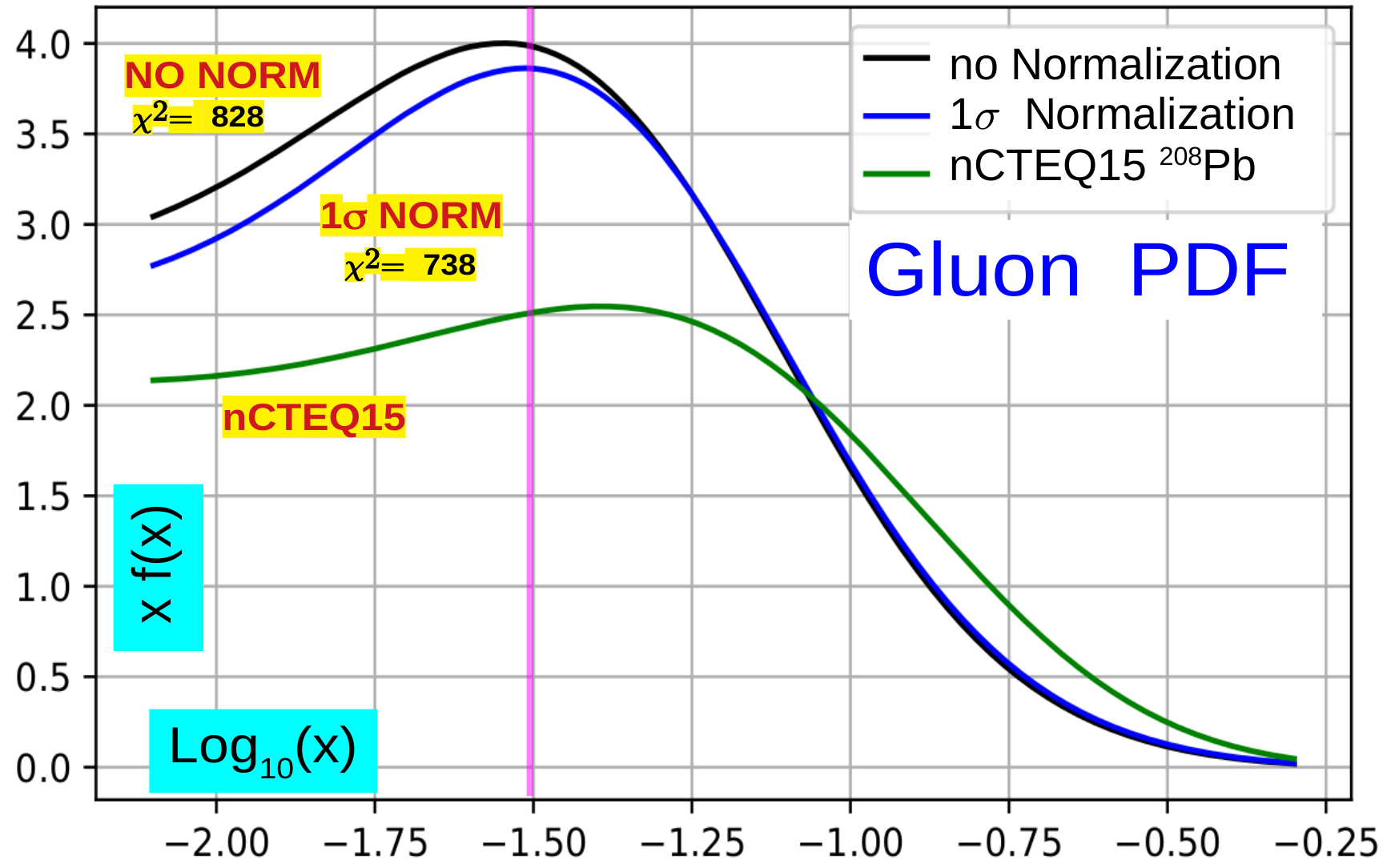}
\vspace{0pt}
\caption{
Resulting nPDFs for lead ($Pb$) at $Q=2$~GeV for the
a)~strange and b)~gluon.
The vertical line (magenta) represents the central $x$ value for pPb \wz{} production. 
}
\label{fig:npdfs}
\end{figure} 
}


\null \vspace{-1.0cm}
\section{Introduction}
\nocite{Kovarik:2015cma}
%
%
The Parton Distribution Functions (PDFs) are the key elements
which allow us to generate concrete predictions for processes with
hadronic initial states.
The success of this theoretical framework has been extensively demonstrated
in fixed-target and collider experiments ({\it e.g.}, at the TeVatron, SLAC, HERA, RHIC, LHC),
and will be essential for making predictions for future facilities (EIC, LHeC, FCC). 
Despite the above achievements, there is yet much to learn about
the hadronic structure and the detailed composition of the
PDFs~\cite{Kovarik:2015cma,Eskola:2016oht,AbdulKhalek:2019mzd}.

Although the up and down PDF flavors are generally well-determined across much of the
partonic $x$ range, there is significant uncertainty in the strange component, $s(x)$.
The strange PDF is especially challenging because, in many processes, it is difficult
to separate this from the down component. 
Fixed-target neutrino--nucleon DIS production of dimuons ($\nu N \to  \mu^+ \mu^- X$)
provided important constraints on $s(x)$; however, as these neutrino experiments were
performed on heavy targets, the nuclear corrections must be considered.\footnote{%
  See Ref.~\cite{Kusina:2012vh} and references therein.}
Proton--proton ($pp$) production of $W$ and $Z$ bosons at the LHC also provides insight on $s(x)$; however, preliminary results
show some tension between the various measurement channels~\cite{Cooper-Sarkar:2018ufj}.

In the current investigation, we will study the
production of $W$ and $Z$ bosons in 
proton--lead ($pPb$) collisions at the LHC;
this involves similar considerations as the $pp$ case,
but also brings in the nuclear corrections. 
We will be focusing, in particular, on the $s(x)$ distribution and
look to compare with the expectations from both fixed-target and $pp$ LHC measurements.

\null \vspace{-1.0cm}
\section{The nCTEQ++ Project}
The  nCTEQ project\footnote{For details, see \href{http://www.ncteq.org}{www.ncteq.org}
  which is hosted at \href{http://www.HepForge.org}{HepForge.org}.} 
extends the  proton PDF global fitting effort by fully including the nuclear dimension. 
Previous to the nCTEQ effort, nuclear data was ``corrected'' to isoscalar data 
and added to the proton PDF fit {\it without} any uncertainties. 
In contrast, the nCTEQ framework allows full communication between the nuclear data 
and the proton data.
%
This enables us to investigate if observed tensions between
data sets could potentially be  attributed to the nuclear corrections.

The details of the nCTEQ program are presented in Ref.~\cite{Kovarik:2015cma}.
The analysis includes   Deeply Inelastic Scattering (DIS), 
lepton pair production (Drell-Yan), and pion production  data
from a variety of experiments totaling 740  data  points (after cuts) and 19 nuclei.
The  computed PDFs compare favorably to other determinations from the 
literature~\cite{Hirai:2007sx,deFlorian:2011fp,Eskola:2016oht}.

 
More recently, the code base was converted to a modular C++
platform (\ncteqpp{}) which  enabled us to easily integrate
programs such as
HOPPET~\cite{Salam:2008qg},
APPLgrid~\cite{Carli:2010rw}, and MCFM~\cite{Campbell:2015qma};  the fit output 
is exported in YAML format  and then processed by Python Jupyter notebooks.
Additionally, using ApplGrids generated from MCFM
we can easily include a wide variety of higher-order processes
directly into the PDF fitting loop. 
An important step in this process was the validation that MCFM
grids were sufficiently ``PDF independent'' so that proton PDF and nuclear PDF grids
could be interchanged.
This groundwork provided the foundation for a nuclear PDF fit including
the NLO \wz{}  production data from the LHC.

\figdy{}
\section{Comparisons: LHC Heavy Ion $\boldmath{W}$ Production with nCTEQ15 PDFs}
\null \vspace{-0.5cm}

In a previous study we compared our predictions
for  the
production of \wz{} bosons
with available LHC data for proton-lead  collisions~\cite{Kusina:2016fxy}.
This process is an ideal QCD ``laboratory'' 
as it  is sensitive to i)~the heavy flavor components $\{s,c,...\}$,
ii)~the nuclear corrections, and iii)~the underlying ``base'' PDF.

In Fig.~\ref{fig:dsigdy} we show selected results of the comparison {\it without  fitting.}
While we found generally good agreement in the negative rapidity region,
the poor agreement  in the positive rapidity region
(which corresponds to small $x$ in the lead PDF) suggests this new
data set can have a significant impact on the resulting PDFs. 
At small $x$, we are in the nuclear ``shadowing'' region
where the lead PDFs are reduced compared to the proton;
if the ``shadowing'' effect were reduced, this would improve
the agreement between data and theory.

In fact, a similar behavior was observed for DIS measurements in
Refs.~\cite{Schienbein:2009kk,Kovarik:2010uv,Owens:2007kp}
when comparing 
$\nu N$ charged-current neutrino DIS processes with 
$\ell^\pm N$ neutral-current DIS processes.
If we use a nuclear correction with a reduced ``shadowing'' correction at small $x$ values,
this would improve both the  $\nu N$ DIS data and the LHC \wz{} comparisons.
Both of these data sets 
are important for distinguishing the various parton flavors---especially the strange PDF.

\figcomp{}

\figchi{}
\section{PDF fit to LHC \boldmath{\wz{}} Data }
We now use the \ncteqpp{} framework to include the NLO LHC  \wz{}  pPb data
into  a PDF fit in addition to  the DIS and DY data sets from the nCTEQ15 fit.
As we are most interested to find the impact of the new data on the
strange quark PDF, we will free up only a limited set of parameters; hence,
this represents only a preliminary fit, and a complete analysis is in progress.\footnote{
  Specifically, we fit 12 parameters: 3 for  $ s+\bar{s} $,
  and the remaining 9 for $\{g, u_V, d_V, \bar{u}+\bar{d} \}$.
This is in contrast to  nCTEQ15 which fits 16 parameters  for $\{g, u_V, d_V, \bar{u}+\bar{d} \}$ and keeps the strange PDF fixed.
}
%
A sample comparison is  displayed in 
Figure~\ref{fig:comp}
where the data for CMS $W^+ $are shown as blue squares with statistical error bars, and the theory in the red circles.
In general, we find the shape of the distributions is well described by our fits;
the normalization issues are more complex.

In Fig.~\ref{fig:chi2} we show the computed $\chi^2/dof$ 
results for the individual experiments.
The separate processes in the figure are color coded.
The DIS data (51xx) is represented by blue bars and
the Drell-Yan (52xx) data by red bars; the fits to these data sets  are generally quite good ($\chi^2\sim 1$).
The  \wz{} data (62xx) is  represented by the green bars.
If we include the  \wz{} data in the fit without allowing for a normalization shift,
the overall $\chi^2/dof$ improves from 992/816 (no fit) to 828/816;
while this is a significant improvement, clearly there are \wz{} data sets with unacceptable $\chi^2/dof$ values.

%
In general, we find that the theory predictions lie below the experimental data; hence, if we allow for a
normalization uncertainty, this additional freedom can significantly improve the fit.
The experiments have an associated luminosity uncertainty, and we will use this as a gauge as we shift the normalizations.
In the second panel of Fig.~\ref{fig:chi2} we show the results allowing for a  normalization shift of up to  $1~\sigma$.
The DIS and DY data (not shown) are essentially unchanged, but this greatly improves the \wz{} fits;
however, there are still a few \wz{} data sets with large  $\chi^2/dof$ values. 
If we allow larger normalization shifts (up to $\sim3\sigma$) for these few data sets, the results are shown in the third panel of  Fig.~\ref{fig:chi2}
and we find it is possible to obtain  $\chi^2/dof\sim 1$ for all data sets. 

While this preliminary exercise demonstrates it is possible to obtain a good fit,
we must ask
i)~how the uncertainties and the normalization issues affect the resulting PDFs,
and
ii)~whether the results truly reflect the underlying physics or are simply an artifact of our fitting procedure. 

We will focus on the the strange and gluon components; these PDFs  show the largest variation, in part, because
they are less constrained than the up and down flavors. 
Fig.~\ref{fig:npdfs} displays the strange and gluon nPDFs for
i)~the original nCTEQ15 set,
ii)~the above fit with no normalization shift [{\small  NO~NORM}],
ii)~the above fit with a $1\sigma$ normalization shift [{\small  $1\sigma$~NORM}].
A vertical line (magenta) indicates the central $x$ value for pPb \wz{} production. 
Compared to the nCTEQ15 result, we see the  [{\small  NO~NORM}] fit pulls the strange PDF up by 40\%
in the $x$ region relevant for \wz{} production.
In contrast, when we allow for a normalization shift [{\small  $1\sigma$~NORM}], the shift
of $s(x)$ is reduced by roughly half.
The above pattern is also reflected in the gluon distribution to a lesser extent.
Thus, the obvious question to ask is the following.
\vspace{-0.1cm}
\begin{quote}
  \textsl{
Are these new data increasing the strange PDF because
that is dictated by nature, or is the fit simply exploiting $s(x)$ because that is one of the least constrained flavors?
}
\end{quote}
\vspace{-0.1cm}
The answer to this important question will require additional study,
and this is currently under investigation with our new nCTEQ++ set of tools.  


\figdouble{}
\section{Conclusion}
\null \vspace{-0.5cm}

Our ability to fully characterize fundamental observables, 
like the Higgs boson couplings and the $W$ boson mass,
and to constrain both SM and BSM signatures is strongly limited by how
accurately we determine the underlying PDFs~\cite{Tanabashi:2018oca}. 
%
A precise determination of the strange PDF is an essential step
in advancing these measurements.

The new nCTEQ++ framework extends the range of processes we are able
to include in our global nPDF analyses.
Specifically,  we were able to include the LHC $W/Z$
data directly in the fit.  While this significantly reduced the
overall $\chi^2$ for the $W/Z$ LHC data, we still observe tensions in particular
data sets which require further investigation.
Our initial analysis has identified factors which might further reduce
the apparent discrepancies observed in the strange quark distribution
including: increasing the strange PDF,
modifying the nuclear correction, and adjusting the data normalization.

The next step is to extend the above preliminary fit with a complete set 
of free parameters and additional data sets  to
help separately disentangle issues of flavor differentiation and nuclear corrections. 
The ultimate goal of the nCTEQ project is to obtain the most precise PDFs using 
the full collection of both proton and nuclear data.


\newpage
 \printbibliography 

\end{document}